\documentclass[10pt]{article}
\usepackage{multirow}
\usepackage[dvips]{graphicx}

\usepackage[psamsfonts]{amssymb}
\usepackage{amsxtra}
\usepackage{threeparttable}
\usepackage{amsmath}
\usepackage{amssymb}

\usepackage{graphicx}

\usepackage{cite}

\usepackage{color}

\usepackage{setspace}
\usepackage{float}

\usepackage[labelfont=bf,labelsep=period,justification=raggedright]{caption}

\makeatletter
\renewcommand{\@biblabel}[1]{\quad#1.}
\makeatother

\date{}

\begin{document}

\begin{flushleft}
{\Large
\textbf{Spatial Auditory BCI Paradigm Utilizing N200 and P300 Responses}
}

Zhenyu Cai$^{1}$, 
Shoji Makino$^{1}$, 
Takeshi Yamada$^{1}$, 
and Tomasz M. Rutkowski$^{1,2,3,\ast}$
\\
\bf{1} Graduate School of Systems and Information Engineering \\
       University of Tsukuba, Japan\\
\bf{2} TARA Center, University of Tsukuba, Tsukuba, Japan\\
\bf{3} RIKEN Brain Science Institute, Wako-shi, Japan\\
$\ast$ E-mail: tomek@tara.tsukuba.ac.jp
\end{flushleft}

\section*{Abstract}

The paper presents our recent results obtained with a new auditory spatial localization based BCI paradigm in which the ERP shape differences at early latencies are employed to enhance the traditional P300 responses in an oddball experimental setting.
The concept relies on the recent results in auditory neuroscience showing a possibility to differentiate early anterior contralateral responses to attended spatial sources. Contemporary stimuli--driven BCI paradigms benefit mostly from the $P300$ ERP latencies in so called ``aha-response'' settings. We show the further enhancement of the classification results in spatial auditory paradigms by incorporating the $N200$ latencies, which differentiate the brain responses to lateral, in relation to the subject head, sound locations in the auditory space. The results reveal that those early spatial auditory ERPs boost online classification results of the BCI application. The online BCI experiments with the multi-command BCI prototype support our research hypothesis with the higher classification results and the improved information-transfer-rates.

\section{Introduction}

The brain computer interface (BCI) utilizes the human neurophysiological signals to control an external computer or a machine~\cite{bciBOOKwolpaw}. BCI does not depend on muscle or peripheral nervous system activities.  Particularly, in case of amyotrophic lateral sclerosis (ALS) suffering patients, it could help them to communicate or to complete various daily tasks (control a computer or type messages on a virtual keyboard, etc). This would create a very good option for ALS patients to communicate with their families, friends, or the caretakers by only using their brain waves. So far, many approaches are focusing on the visual modality BCI applications. The visual modality BCI can not be used by ALS patients who often suffer in the advanced disease stages from limited or lost sight. We present in this paper a concept of an auditory BCI based on spatial sound stimuli, which we call shortly saBCI (spatial auditory BCI). The saBCI concept is based on a basic feature of the human auditory pathway which is very sensitive to localization of changing spatial auditory sources~\cite{book:auditoryNEUROSCIENCE}. The auditory pathway has also very good temporal resolution, which is an additional feature we would like to utilize in the saBCI design. This will allow to decrease stimuli onset asynchrony (SOA) of the presented sound stimuli in comparison to vision based applications~\cite{iwpash2009tomek}.

The contemporary stimuli--driven BCI approaches are mostly based on P300 responses to distinguish \emph{targets} and \emph{non--targets} from series of event related potential (ERP) responses~\cite{bciSPATIALaudio2010}. Recently a new result~\cite{spatialAUDIOerp2011} was published elucidating ``the $N200$-anterior-contralateral'' ($N2ac$) component at the early latency (around $200$ms) of an auditory ERP. The $N2ac$ was obtained in an experiment using two $750$ms long sound stimuli which were presented simultaneously from a different loudspeaker each. Subjects were requested to attended to the instructed \emph{target} sound that could occur  from any loudspeaker.

\begin{figure}[H]
	\begin{center}
		\includegraphics[width=0.6\linewidth]{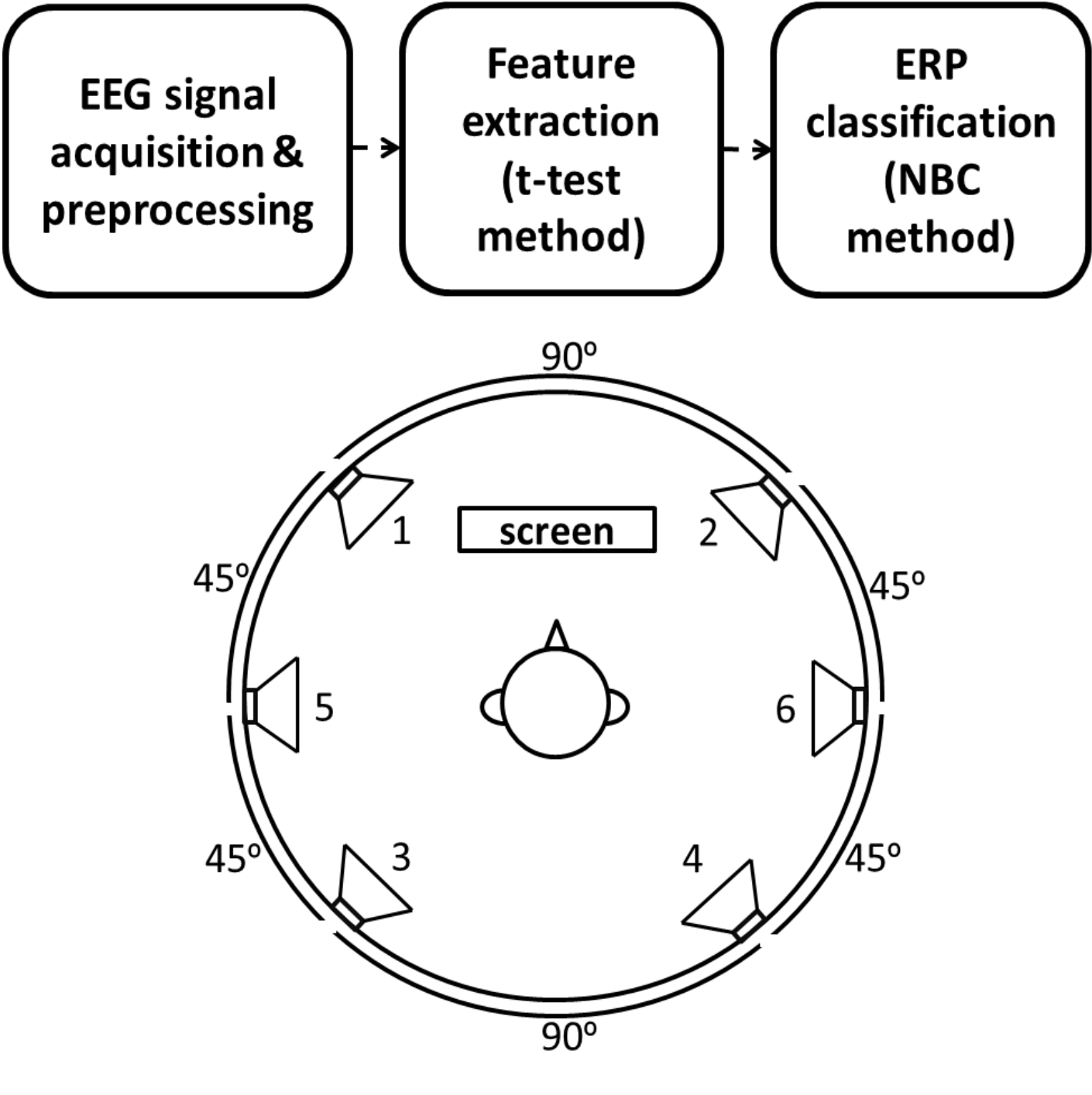}
	\end{center}
	\caption{The novel $N2apc$ paradigm based on spatial sound stimuli}\label{fig:spatialAUDIOstimuliEEG_6sp}
\end{figure}

We propose to design the new saBCI experimental paradigm based on the auditory spatial localization principle as the informative cues with support of the  $N2ac$ component elicited in the new setup as depicted in Figure~\ref{fig:spatialAUDIOstimuliEEG_6sp}. Our hypothesis is that the new ERP component shall improve the classification results and the final information transfer rate (ITR) leading to a better BCI usage comfort.

Within the novel saBCI paradigm framework, the subjects are asked, as in usual oddball paradigm, to attend  and count the target stimuli from the instructed or intended direction, while ignoring the other. The EEG signals are recorded with \textsf{g.MOBIlab+} EEG amplifier by \textsf{g.tec}. We use the novel dry  \textsf{g.SAHARA} electrodes by the same producer which further improve the interfacing comfort, since there is no need to apply a conductive gel.
In order to decrease the unnecessary and signal quality degrading muscular movement related electromyography (EMG) noise on ERP responses, the subjects are asked to minimize their eye, facial and body in general movements during the experiments.

In our previous paper~\cite{tomekAPSIPA2011} we proposed a channel and ERP--latency selection in order to improve the classification results. At that time the $P300$ response (so called ``aha--response'' at the latency around $300$ms elicited to the expected/instructed target stimulus~\cite{book:eeg}) was the major feature used for the classification of the attended \emph{targets} of the oddball paradigm. In this paper, we introduce the early latencies around $200$ms ($N200$ response) which precede the $P300$. They shall improve the final classification rates of the saBCI application.

The objective of this paper is to test and confirm our working hypothesis that the auditory evoked response based on $N2ac$ paradigm should improve the saBCI application classification rates based on the new lateral to the subject head stimuli responses analysis.

From now on the paper is organized as follows. In the next section the experimental setup and the novel paradigm is described together with EEG signals pre-processing steps. Next the analysis and optimization procedures of the ERPs at $N200$ and $P300$ response latencies for all experimental subjects are described. Finally classification and ITR results discussion conclude the paper together with future research directions.

\section{Methods}

The EEG experiments to validate the proposed spatial auditory BCI paradigm utilizing the $N200$ and $P300$ latency responses have been conducted in Multimedia Laboratory in TARA Life Science Center at the University of Tsukuba, Tsukuba, Japan.
All the experimental procedure details and this approach research targets have been explained to the seven human subjects who agreed voluntarily to attend.
The experimental procedures are designed in accordance with ethical committee guidelines of this paper author affiliated institutions.
The EEG signals are recorded by the \textsf{g.USBamp} EEG amplifier with the six dry \textsf{g.SAHARA} electrodes. The sampling frequency is set to $256$Hz with a notch filter to reject the $50$Hz power--line noise.

The auditory stimuli has been presented through six loudspeakers distributed with an equal radius of $1$ meter around the subject's head as depicted in Figure~\ref{fig:spatialAUDIOstimuliEEG_6sp}. Three speakers with equal distances are positioned at each lateral side to the head. Two short \emph{white}-- and \emph{pink}--noise stimulus bursts are used as described in the following section.
All the experiments are conducted in a silent and low reverberation room in order to limit an interference of ``an environmental noise.''

\subsection{The Offline saBCI Experimental Protocol}

The experimental hypothesis is that we shall be able to distinguish from the ERP shape which direction (left or right) the subject attends based on the novel $N2ac$ response analysis method.

To test the hypothesis we conduct a series of EEG recording experiments in the offline BCI mode~\cite{bciBOOKwolpaw} (no instant feedback or classification results given to the subject). The experiments are performed with the seven healthy subjects (six males and one female; age range $21-42$ with the mean of $26.4$ years old). The experimental procedure has been explained in detail to each subject and her/his consent has been obtained.
The subject is seated in the center of the experimental studio and the dry EEG electrodes are attached on the scalp.
The subject's chair position is surrounded by the six loudspeakers. The elevation of the loudspeakers is fixed to the subject's ear level. A computer display with experimental instruction is set in front of the subject. The six loudspeakers are distributed on a circle with the three loudspeakers $(1,3,5)$ positioned on the left side with $45$--degrees angular distance. The remaining three loudspeakers $(2,4,6)$ are located on the right side with the same angular distances (see Figure~\ref{fig:spatialAUDIOstimuliEEG_6sp}).

The sound stimulus is presented in random order one at a time from a single loudspeaker (a single trial consists of a delivery of a single \emph{target} and five \emph{non-targets}). We decide to use two broadband noise stimulus types in order to utilize two spatial localization mechanism of the human auditory pathway (the inter-aural time and level differences - ITD/ILD)~\cite{book:auditoryNEUROSCIENCE}. The \emph{white}-- and \emph{pink}--noise stimuli of $30$ms lengths with $5ms$ linear attack and sustain periods has been chosen. The SOA is set to $300$ms.
The single session consists of the six single trials ($6$ \emph{targets} from each direction accompanied by $30$ \emph{non--targets}). The \emph{target} direction in each trial is presented randomly together with five \emph{non--targets}.
For each subject and each stimuli we perform $15$ sessions (all together $90$ \emph{targets} and $450$ \emph{non-targets} are delivered). The target direction instruction is presented visually on a computer display and auditory from the same loudspeaker which subject shall latter attend to. Before each experiment the subjects are allowed for a short practice session to familiarize themselves with spatial auditory conditions.

\section{The Analysis of ERP Responses in Offline BCI Paradigm}

In many current auditory BCI applications the focus is put on a binary classification of brain evoked responses to \emph{targets} versus \emph{non--targets}~\cite{bciSPATIALaudio2010,tomekAPSIPA2011,spatialDYNBCIfrontiersNEURO2011,spatialBCI9CLASSfrontiersNEURO2011}. The majority of the contemporary BCI applications aim at the $P300$ response latency without consideration of the remaining ERP ranges. Only a single of recently published papers mentions the $N200$ latency range as possibly useful to support classification~\cite{spatialBCI9CLASSfrontiersNEURO2011}, but there is no comparison made so far with $P300$ only related results, what we attempt in this paper.  We compare and discuss the $N200$ response suitability and we show that it really improves the classification results.

Basically a concept of adding the early latency $N2ac$ response is based on our previous~\cite{tomekAPSIPA2011} research and the recently published by other groups~\cite{spatialAUDIOerp2011} concept of this ERP range modulation by \emph{ipsilateral} vs. \emph{contralateral} stimulus spatial locations. The \emph{ipsilateral} $N2ac$ response has higher amplitude comparing to the \emph{contralateral} one. This difference confirms a feasibility to utilize the early $N200$ response latency  to improve the \emph{target} vs. \emph{non--target} classifications outcomes.

In order to precisely analyze an impact of the early ERP reposes on the saBCI paradigm classification we propose to conduct two separate analyses that shall compare how much the improvement depends only on the $N200$ response feature addition, and how much on the new feature composition based on the comparison of the \emph{ipsilateral} and \emph{contralateral} responses as in $N2ac$ design.

\subsection{EEG Preprocessing}

The EEG signals captured by the  \textsf{g.MOBIlab+} system with \textsf{g.SAHARA} dry electrodes are first filtered digitally with the two $5^{th}$--order Butterworth high-- and low--pass filters with cutoff frequencies at $0.5$Hz and $25$Hz, respectively.

The high--pass filtering removes the very slow baseline drift related artifacts as well as the slow eye movements related EMG interferences. The low--pass filter limits the higher frequency EMG artifacts related to subject body muscle movements.

Next the EEG signals are segmented creating the ERP related \emph{epochs}. Each \emph{epoch} starts $100$ms before each stimuli onset and ends $700$ms after it. We use the $100$ms pre--stimuli onset interval for a baseline correction procedures.

In the next step the eye movement artifacts rejection is carried out. Auditory spatial stimuli are known to cause in subjects the uncontrolled eye movements~\cite{tomekHAID2011} which in the current approach are removed with a threshold value set at $80\mu$V (signal amplitude level above the usual EEG activity). The rejected epochs are not further processed, since in the current approach an emphasis is focused on the spatial paradigm validation.

\subsection{The Optimization of the EEG Electrode Locations and ERP Features Extraction}

\begin{figure}[H]
	\begin{center}
		\includegraphics[width=0.7\linewidth]{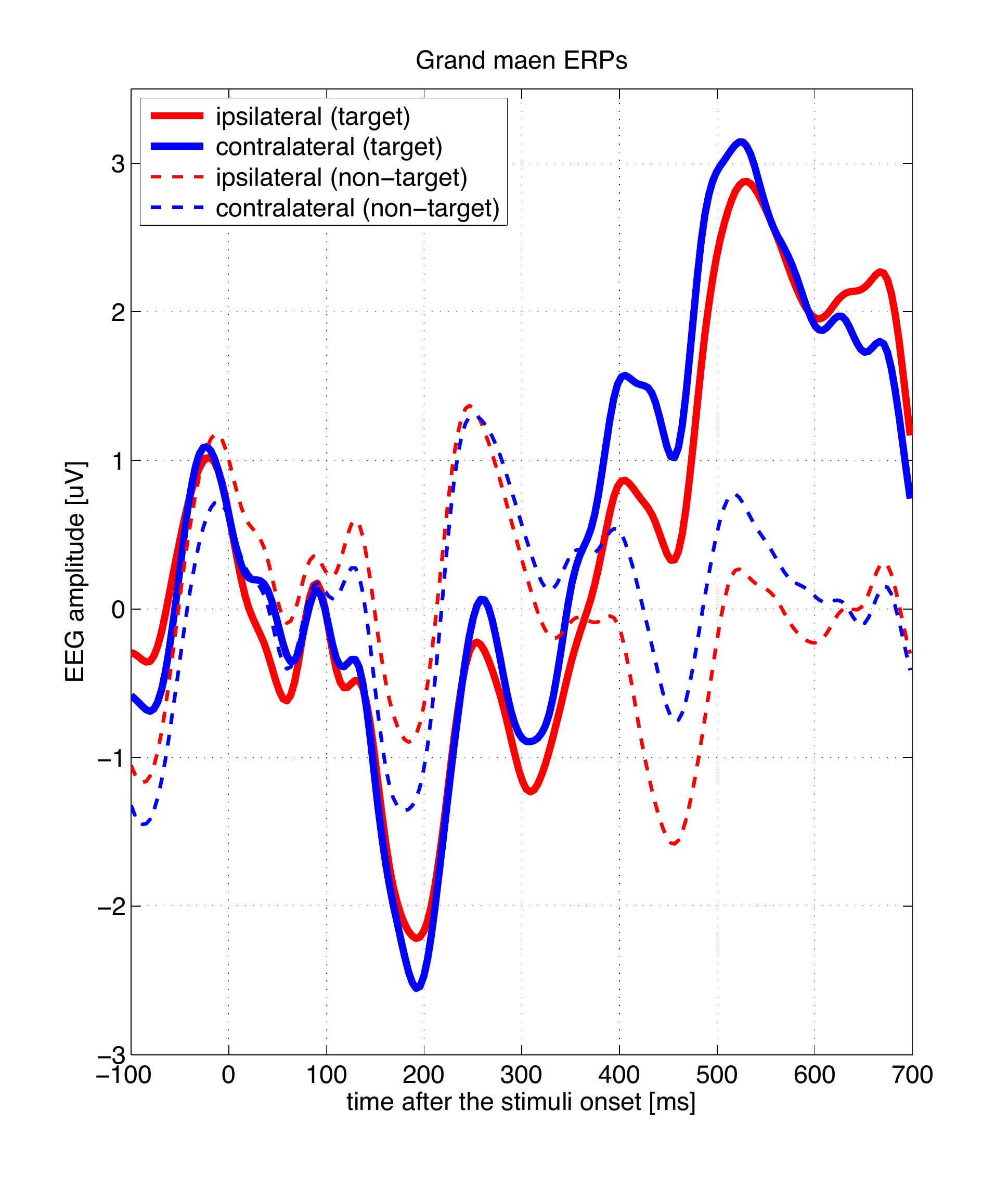}
	\end{center}
	\caption{The grand mean averaged ERP responses of the seven subjects. The solid lines depict \emph{targets} and the dashed ones \emph{non--targets}. The red color indicates ipsilateral and blue one the contralateral responses. The differences between \emph{targets} and \emph{non--targets} are obvious after $300$ms (the so called ``aha''-- or $P300$ response), while the lateral directions can be identified in $N200$ latency area.}\label{fig:GrandMeanN2acResults}
\end{figure}

In the previously reported research on $N2ac$ phenomenon~\cite{spatialAUDIOerp2011} the anterior cluster of electrodes sites $F3$, $F7$, $C3$, $T7$, $F4$, $F8$, $C4$, and $T8$  was used, as in $10/20$--\emph{international system}~\cite{Jurcak20071600}. In our experimental setup, we select the $F5$, $F6$, $C3$, $C4$, $P5$, and $P6$ electrodes in order to have additional responses from parietal cortices known to generate ERPs related to spatial and $P300$ responses~\cite{book:eeg}. Additionally we show that the $P5$ and $P6$ sites are also useful to differentiate the responses to lateral stimuli similarly as for left--right only comparison revealed by $N2ac$. We call the new finding the $N2apc$ ($N200$--anterior--posterior--contralateral) as extension of the former one.
An example in Figure~\ref{fig:GrandMeanN2acResults} shows the averaged and artifact--removed classical $N2ac$ responses to \emph{ipsilateral} and \emph{contralateral} sound stimuli as confirmed by our experiments. The presented $N200$ area responses are elucidated for \emph{ipsilateral} and \emph{contralateral} targets.

\begin{figure}[H]
	\begin{center}
	\vspace{-0.35cm}
		\includegraphics[width=0.7\linewidth]{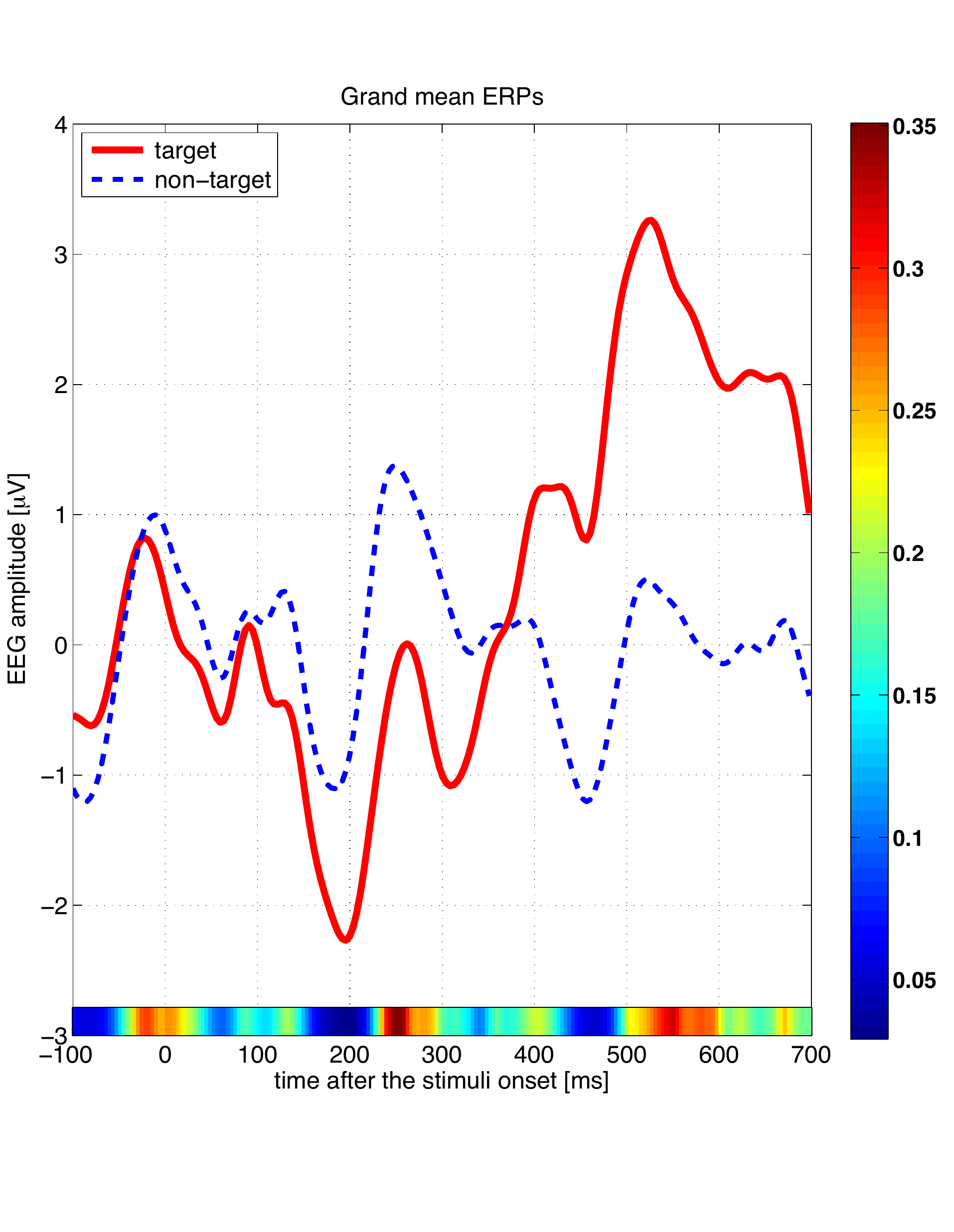}
	\end{center}
	\vspace{-0.7cm}
	\caption{The grand mean averaged ERP for the all seven subjects and all electrodes calculated together, while plotted separately for \emph{target} (solid red line) and  \emph{non--target} (dashed blue line) responses. The significant differences between the both responses can be found, as visualized by the color bar with $p$-values of \emph{t--test} results (statistical significance for $p<0.05$) in the bottom part in the above panel, can be found around $200$ms ($N200$ response latency) and after $300$ms ($P300$ response latency).}\label{fig:grandmeanTargetNontargetRestults}
\end{figure}

In order to validate statistically the differences between \emph{target} and \emph{non--target} responses we conduct the \emph{t--test} analysis of the two class ERP means~\cite{bookTERRbio} in \emph{ipsilateral} vs. \emph{contralateral} experimental setting.
The \emph{t--test} method is applied to compare the differences of response distributions in single trials for each sample point of the collected ERPs. As the result we can  extract discriminative information (in $N200$ and $P300$ latencies) leading to later classification optimization. The results of the above analysis are depicted in Figure~\ref{fig:grandmeanTargetNontargetRestults}. A color bar located on a time scale in the above figure visualizes the \emph{t--test's} $p$ value results, which is a probability of the null hypothesis rejections that the means from the both compared distributions are significantly different (usually $p<0.05$ in life sciences is considered as the significant value). The color bar in the Figure~\ref{fig:grandmeanTargetNontargetRestults} clearly shows that the postulated $N2apc$ differential response for lateral responses is located in the range from $100$ms to $300$ms, similarly to the previously published $N2ac$ one. This finding confirms our hypothesis, that the early $N200$--range latencies are related to spatial localization processes in the human brain and that the parietal electrodes contribute also to the result.

In this paper two types of binary classification problems are discussed. First we evaluate our first hypothesis that adding the early latency ERP periods as features improves the binary \emph{target} vs. \emph{non-target} classification. Next we also show that the novel $N2apc$ response further enhances the results using the \emph{ipsilateral} vs. \emph{contralateral} response comparison.

\begin{figure}[H]
	\begin{center}
		\includegraphics[height=18cm]{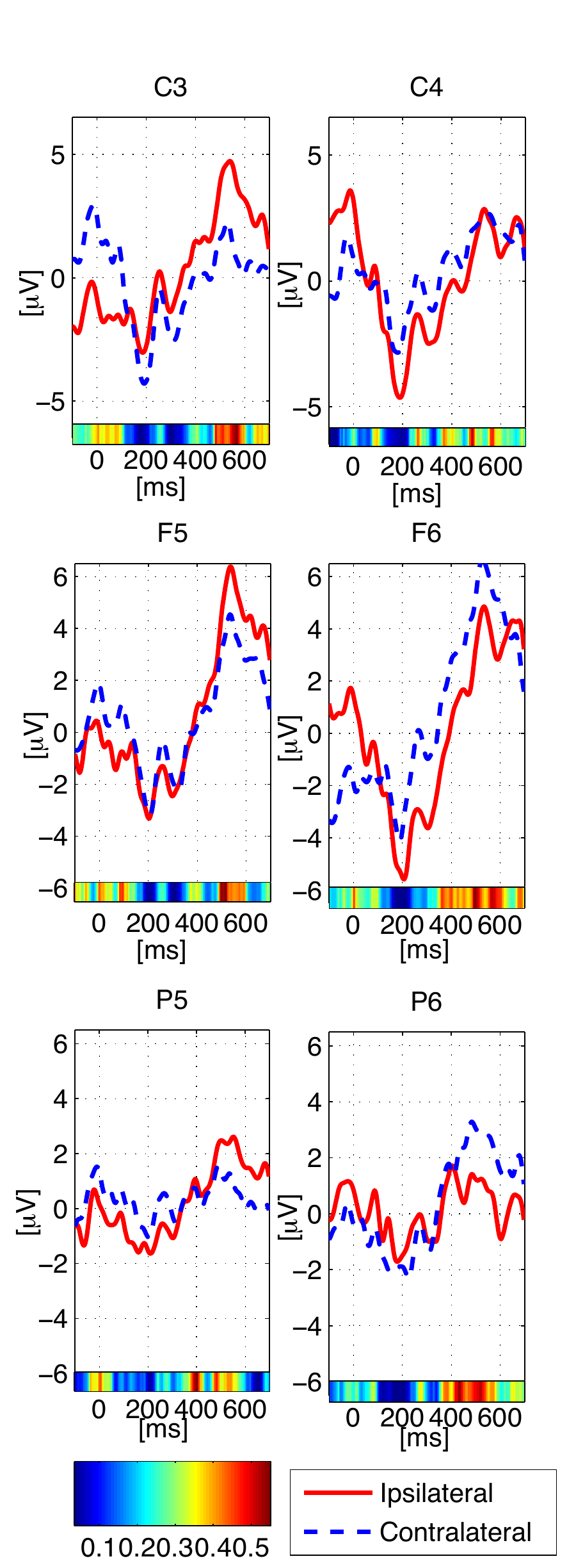}
	\end{center}
	\caption{ERP to pink noise stimuli grand mean averages  for all subjects and the six electrodes plotted separately in each panel. The solid red lines represent the \emph{ipsilateral} to target responses and the dashed blue lines to the\emph{contralateral} ones. The color bars at the bottom of each panel show the \emph{t--test} resulting $p$-values.}\label{fig:Ttest6chN2acPINKNOISE}
\end{figure}

\begin{figure}[H]
	\begin{center}
		\includegraphics[height=18cm]{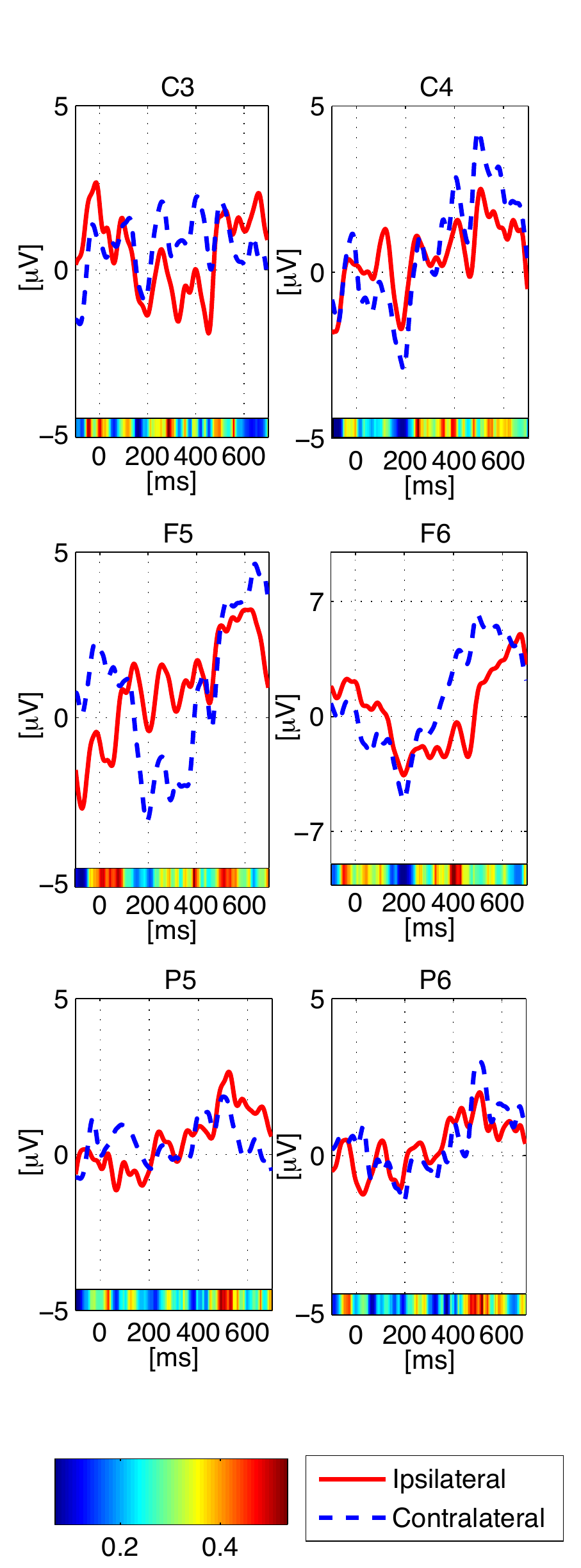}
	\end{center}
	\caption{ERP to white noise stimuli grand mean averages  for all subjects and the six electrodes plotted separately in each panel. The solid red lines represent the \emph{ipsilateral} to target responses and the dashed blue lines to the\emph{contralateral} ones. The color bars at the bottom of each panel show the \emph{t--test} resulting $p$-values.}\label{fig:Ttest6chN2acWHITENOISE}
\end{figure}

In order to find the most discriminable features from ERP responses we use the results of the above described \emph{t--tests} evaluating statistical significance of them. We ``hand pick'' only those samples within each subject's ERPs for which the $p$--values are smaller than $0.05$ as depicted by blue shades of the color bar at the bottom of the Figure~\ref{fig:grandmeanTargetNontargetRestults}.
The significantly different ERP samples of $N2apc$ based experiments (we relax here the condition to $p<0.10$ only) are depicted in Figures~\ref{fig:Ttest6chN2acPINKNOISE} and ~\ref{fig:Ttest6chN2acWHITENOISE} for pink-- and white--noise stimuli respectively. In the next section we show that the relaxed condition of \emph{t--test's} $p<0.10$ improves already satisfactory the saBCI classification results by incorporating the $N200$ latency responses.

\subsection{The Offline saBCI Classification}\label{sec:NBC}

We perform the classification steps for each subject separately in saBCI offline mode, which means that all procedures are conducted after each experiment of data collection, without any online feedback to subjects.
The classification procedure is performed in a so called binary task paradigm (we classify \emph{target} vs. \emph{non--target}, or \emph{contralateral} vs. \emph{ipsilateral} response pairs each time only).

In each classifier training and testing step we select $90$ \emph{targets} and a random subset of $90$ \emph{non-targets} (from the $450$ available) to have the balanced number of the members in each class set. The resulting chance level is $50\%$.
For the case of the \emph{contralateral} vs. \emph{ipsilateral} responses classification we select $30$ \emph{contralateral} and $30$ \emph{ipsilateral} events.

Based on our previous classification trials reported in~\cite{tomekAPSIPA2011} we decide to use a Baysian classifier, which outperforms the linear discrimination analysis methods. \emph{The naive-Bayses classifier} (NBC) is particularly suited for the highly dimensional features.
Despite its simplicity, the NBC approach often outperforms more sophisticated classification methods~\cite{book:pattRECO}. The NBC application assigns an unknown sample (ERP features in our case) $\mathbf{x} = [x_1,x_2,\ldots,x_l]^T$ based on probability maximization to the class
\begin{equation}
	\omega_m = \arg\max_{\omega_i}\prod_{j=1}^l p(x_j|\omega_i), \quad i=1,2,\ldots,M, \label{eq:NBCclass}
\end{equation}
with an assumption that the individual features $x_j$, $j=1,2,\ldots,l$, shall be statistically independent. It turns out that the NBC can be very robust also to violations of the independence assumption~\cite{book:pattRECO}. 

Consider the vector $\mathbf{x}$ with features according to the values of the ERP ``hand picked'' samples. The respective conditional probabilities shall be $P(x_i|\omega_1) = p_i$ and $P(x_i|\omega_2) = q_i$, in our binary classification case comparing \emph{targets} vs. \emph{non--targets} or \emph{ipsilateral} vs. \emph{contralateral} responses. In Bayesian rule, given the value of $\mathbf{x}$ the class membership is decided according to the probabilities likelihood ratio
\begin{equation}
	\frac{P(\omega_1)P(\mathbf{x}|\omega_1)}{P(\omega_2)P(\mathbf{x}|\omega_2)} > (<) 1. \label{eq:NBCclassPROB}
\end{equation}
The adoption of features independence principle allows us to limit a number of necessary training samples and we can write
\begin{eqnarray}
	P(\mathbf{x}|\omega_1) = \prod_{i=1}^l p_i^{x_i}(1-p_i)^{1-x_i}\\
	P(\mathbf{x}|\omega_2) = \prod_{i=1}^l q_i^{x_i}(1-q_i)^{1-x_i}
\end{eqnarray}
Now an application of a logarithm function to the both sides of the equation~(\ref{eq:NBCclassPROB}) results with a linear discriminant function as
\begin{eqnarray}
	h(\mathbf{x}) &=& \sum_{i=1}^l \left(x_i \ln\frac{p_i}{q_i}+(1-x_i)\ln\frac{1-p_i}{1-q_i}\right)\\ \nonumber
	&+& \ln\frac{P(\omega_1)}{P(\omega_2)},
\end{eqnarray}
which could be brought to the linear form of
\begin{equation}
	h(\mathbf{x}) = \mathbf{w}^T \mathbf{x} + w_0,
\end{equation}
based on the following substitutions
\begin{eqnarray*}
	\mathbf{w} &=& \left[\ln\frac{p_1(1-q_1)}{q_1(1-p_1)},\ldots,\ln\frac{p_l(1-q_l)}{q_l(1-p_l)}\right]^T\\
	w_0 &=& \sum_{i=1}^l \ln\frac{1-p_i}{1-q_i}+\ln\frac{P(\omega_1)}{P(\omega_2)}.
\end{eqnarray*}
The results of NBC technique successful application are presented in the next
section.

\section{Results}

As the result of the presented research have obtained the results showing that for the both experimental settings of saBCI offline paradigm the classical $P300$ latency could be improved with the pure $N200$ or the more complex $N2apc$ features identified with $p$--values calculated using the classical \emph{t--test} for significance. We summarize below the obtained results.

\subsection{The Classification Results from the Combined $N200$ and $P300$ ERP Latencies in the Classical \emph{target} vs. \emph{non--target} Setting}\label{sec:N200P300class}

The first summary of classification results is presented in Table~\ref{tab:classRESULTSone}, where classification accuracies for the features drawn from $N200$, $P300$ and the combined latencies are shown. The majority of the subjects performed already above the chance level of $50\%$ (except subject MA for the pink noise case) for single feature latencies of $N200$ or $P300$. The proposed combination of the two ``hand--picked'' feature sets using the \emph{t--test} significant ERP samples allowed us to boost the classification results up to $7\%$ (only a single case of the accuracy decrease has been reported) using the \emph{leave--one--out} cross validation~\cite{book:pattRECO} for the NBC technique.

\subsection{The Classification Results from the new $N2apc$ ERP Feature in the Ipsilateral vs. Contralateral Settings}

The results of the proposed approach to compare \emph{ipsilateral} and \emph{contralateral} to target evoked potentials have been summarized in the Table~\ref{tab:classRESULTStwo}, based on the ERP features drawn from results of the \emph{t--test} analysis as summarized in the Figures~\ref{fig:Ttest6chN2acPINKNOISE}~and~\ref{fig:Ttest6chN2acWHITENOISE}. The classification accuracy results have been $17\%$ boosted in the best case, with the same method of the NBC \emph{leave--one--out} cross validation.

\subsection{Analysis of Information Transfer Rate Improvement Results}

The amount of information carried by every selection in the BCI application is usually quantified by the ITR which is calculated based on bits--per--selection $R$, defined as in~\cite{bciSPATIALaudio2010}:
\begin{equation}
	R = \log_2 N + C\cdot\log_2 C + (1-C)\cdot\log_2\left(\frac{1-C}{N-1}\right),\label{eq:bitRATE}
\end{equation}
where $C$ is the classification accuracy and $N$ is the number of classes ($N=6$ in this paper). The final \emph{bit--per--minute--rate} $B$ is obtained after a multiplication by a classification speed $V$, resulting in selections/minute [bit/min] as:
 \begin{equation}
	B = V \cdot R\label{eq:ITR}
\end{equation}
The ITR results are summarized in Tables~\ref{tab:ITRp300n200}~and~\ref{tab:ITRn2apc}. For the both cases of the $N200/P300$ combination and the $N2apc$ paradigm, there is a significant increase of ITR for the majority of subjects.

\section{Conclusions}

In this paper we presented two approaches leading to improvements of classification accuracy and ITR in offline saBCI paradigm by introducing the novel ERP feature extraction in combined $N200/P300$ latencies and in the new $N2apc$ setting which compares responses of lateral, to the head, sound sources.

The first improvement analysis resulted in a comparison of classification rates for the three ERP feature sets of $N200$ and $P300$ latencies processed separately, versus the combined $N200/P300$. The latter combination resulted in a steady increase in classification accuracy for the majority of subjects up to $7\%$ at maximum. Additionally the ITR improvement in this case was reported at maximum of $7$bit/min. This is a very good result giving a possibility to further improve the auditory paradigm based BCI.

The second improvement step is based on the proposed extension of $N2ac$ concept. We added a comparison of parietal electrodes responses allowing for the new feature creation from such ERP comparisons. The new ERP component was named $N2apc$ since it combines anterior and posterior contralateral response differences. The obtained classification and ITR improvement was also very encouraging.

The two main achievements reported in the paper allowed us to improve the novel saBCI paradigm in offline mode which is a step forward in the non--vision based interfacing strategies.
The obtained results reveal that not only the cortical auditory information processing centers related to the cognitive streams could be utilized to BCI purposes. Also the differences in ERPs at early latencies before $300$ms are useful and they guarantee good classification results and ITRs. These results reveal that the very early spatial auditory ERPs are potentially interesting for faster BCI applications.

\section*{Acknowledgements}

This research was supported in part by the Strategic Information and Communications R\&D Promotion Programme no. 121803027 of The Ministry of Internal Affairs and Communication in Japan, and by KAKENHI, the Japan Society for the Promotion of Science grant no. 12010738.

The first author's conference trip was supported by Tateisi Science and Technology Foundation in Japan, grant no. 2022105.

\bibliographystyle{ieeetr}
\bibliography{cai}

\newpage
\section*{Figure Legends}

\begin{description}
	\item[Figure~\ref{fig:spatialAUDIOstimuliEEG_6sp}] The novel $N2apc$ paradigm based on spatial sound stimuli.
	\item[Figure~\ref{fig:GrandMeanN2acResults}.] The grand mean averaged ERP responses of the seven subjects. The solid lines depict \emph{targets} and the dashed ones \emph{non--targets}. The red color indicates ipsilateral and blue one the contralateral responses. The differences between \emph{targets} and \emph{non--targets} are obvious after $300$ms (the so called ``aha''-- or $P300$ response), while the lateral directions can be identified in $N200$ latency area.
	\item[Figure~\ref{fig:grandmeanTargetNontargetRestults}] The grand mean averaged ERP for the all seven subjects and all electrodes calculated together, while plotted separately for \emph{target} (solid red line) and  \emph{non--target} (dashed blue line) responses. The significant differences between the both responses can be found, as visualized by the color bar with $p$-values of \emph{t--test} results (statistical significance for $p<0.05$) in the bottom part in the above panel, can be found around $200$ms ($N200$ response latency) and after $300$ms ($P300$ response latency).
	\item[Figure~\ref{fig:Ttest6chN2acPINKNOISE}] ERP to pink noise stimuli grand mean averages  for all subjects and the six electrodes plotted separately in each panel. The solid red lines represent the \emph{ipsilateral} to target responses and the dashed blue lines to the\emph{contralateral} ones. The color bars at the bottom of each panel show the \emph{t--test} resulting $p$-values.
	\item[Figure~\ref{fig:Ttest6chN2acWHITENOISE}] ERP to white noise stimuli grand mean averages  for all subjects and the six electrodes plotted separately in each panel. The solid red lines represent the \emph{ipsilateral} to target responses and the dashed blue lines to the\emph{contralateral} ones. The color bars at the bottom of each panel show the \emph{t--test} resulting $p$-values.
\end{description}

\newpage
\section*{Tables}

\begin{table}[H]
\begin{center}
\begin{threeparttable}
\caption{The classification results for ERP latencies in $N200$ and $P300$ responses for \emph{target} vs. \emph{non--target} paradigm. The three feature sets ($N200$, $P300$ and $N200/P300$ latencies combined) classification results are compared. The classification improvement comparing the classical $P300$ latency only with the proposed combination of $N200/P300$) is summarized in the right column.}\label{tab:classRESULTSone}
\begin{tabular}{|c|c|c|c|c|c|}
\hline
& noise & N200 & P300 & N200/P300 & N200/P300\\
subject & stimulus & only & only & combined & vs. P300\\
& type & $[\%]$ & $[\%]$ & $[\%]$ & $[\%]$\\
\hline\hline
\multirow{2}{*}{ZH} & pink & $63$ & $63$ & $64$ & $1$\\
& white & $54$ & $59$ & $60$  & $1$ \\
\hline
\multirow{2}{*}{TO} & pink & $52$ & $54$ & $56$ & $2$\\
& white & $56$ & $69$ & $68$ & $-1$\\
\hline
\multirow{2}{*}{NI} & pink & $53$ & $57$ & $57$ & $0$\\
& white & $57$ & $57$ & $58$ & $1$\\
\hline
\multirow{2}{*}{MO} & pink & $60$ & $69$ & $69$ & $0$\\
& white & $55$ & $58$ & $65$  & $7$\\
\hline
\multirow{2}{*}{MA} & pink & $65$ & $65$ & $67$ & $2$\\
& white & $46$ & $40$ & $44$ & $4$\\
\hline
\multirow{2}{*}{CH} & pink & $54$ & $59$ & $59$ & $0$\\
& white & $53$ & $52$ & $53$ & $1$\\
\hline
\multirow{2}{*}{CA} & pink & $64$ & $61$ & $66$ & $5$\\
& white & $53$ & $61$ & $63$ & $2$\\
\hline
\end{tabular}
\end{threeparttable}
\end{center}
\end{table}

\begin{table}[H]
\begin{center}
\begin{threeparttable}
\caption{The classification results for the proposed method using $N2apc$ response to support the saBCI compared with the conventional method.}\label{tab:classRESULTStwo}
\begin{tabular}{|c|c|c|c|c|c|}
\hline
& noise & conventional & N2apc & the\\
subject & stimulus & method & paradigm & improvement \\
& type & $[\%]$ & $[\%]$ & $[\%]$ \\
\hline\hline
\multirow{2}{*}{ZH} & pink & $56$ & $61$ & $5$\\
& white & $51$ & $63$ & $12$ \\
\hline
\multirow{2}{*}{TO} & pink & $52$ & $61$ & $9$\\
& white & $63$ & $67$ & $4$\\
\hline
\multirow{2}{*}{NI} & pink & $58$ & $58$ & $0$\\
& white & $47$ & $54$ & $7$\\
\hline
\multirow{2}{*}{MO} & pink & $37$ & $54$ & $17$\\
& white & $49$ & $50$ & $1$\\
\hline
\multirow{2}{*}{MA} & pink & $49$ & $54$ & $5$\\
& white & $50$ & $56$ & $6$\\
\hline
\multirow{2}{*}{CH} & pink & $32$ & $48$ & $16$\\
& white & $48$ & $50$ & $2$\\
\hline
\multirow{2}{*}{CA} & pink & $72$ & $68$ & $-4$\\
& white & $58$ & $69$ & $11$\\
\hline
\end{tabular}
\end{threeparttable}
\end{center}
\end{table}

\begin{table}[H]
\begin{center}
\begin{threeparttable}
\caption{The ITR, see equations~(\ref{eq:bitRATE})~and~(\ref{eq:ITR}), for the three ERP interval related classification approaches using $N200$ or $P300$ only, and the combined $N200/P300$ together.}\label{tab:ITRp300n200}
\begin{tabular}{|c|c|c|c|c|c|}
\hline
& noise & N200 & P300 & N200/P300 & N200/P300\\
subject & stimulus & only & only & combined & vs. P300\\
& type & [bit/min] & [bit/min] & [bit/min] & [bit/min]\\
\hline\hline
\multirow{2}{*}{ZH} & pink & $25.84$ & $25.84$ & $26.88$ & $1.04$\\
& white & $17.38$ & $21.88$ & $22.84$  & $0.96$ \\
\hline
\multirow{2}{*}{TO} & pink & $15.72$ & $17.38$ & $19.12$ & $1.74$\\
& white & $19.12$ & $32.40$ & $31.25$ & $-1.15$\\
\hline
\multirow{2}{*}{NI} & pink & $16.14$ & $20.02$ & $20.02$ & $0.00$\\
& white & $22.84$ & $20.02$ & $20.94$ & $0.92$\\
\hline
\multirow{2}{*}{MO} & pink & $22.84$ & $32.40$ & $32.40$ & $0.00$\\
& white & $18.24$ & $20.94$ & $27.94$ & $7.00$\\
\hline
\multirow{2}{*}{MA} & pink & $27.94$ & $27.94$ & $30.13$ & $2.19$\\
& white & $11.19$ & $7.36$ & $9.84$  & $2.48$\\
\hline
\multirow{2}{*}{CH} & pink & $17.38$ & $21.88$ & $21.88$ & $0.00$\\
& white & $16.54$ & $15.72$ & $16.54$  & $0.82$\\
\hline
\multirow{2}{*}{CA} & pink & $26.88$ & $23.82$ & $29.02$ & $5.20$\\
& white & $16.54$ & $23.82$ & $25.84$ & $2.02$\\
\hline
\end{tabular}
\end{threeparttable}
\end{center}
\end{table}

\begin{table}[H]
\begin{center}
\begin{threeparttable}
\caption{The ITR, see equations~(\ref{eq:bitRATE})~and~(\ref{eq:ITR}), for the proposed method using $N2apc$ response to support the saBCI classification rates.}\label{tab:ITRn2apc}
\begin{tabular}{|c|c|c|c|c|}
\hline
 & noise & conventional & proposed & resulting \\
subject & stimulus & method & N2apc & change \\
& type & [bit/min] & [bit/min] & [bit/min] \\
\hline\hline
\multirow{2}{*}{ZH} & pink & $19.12$ & $23.82$ Ê& $4.70$\\
& white & $14.92$ & $25.84$ Ê& $10.92$\\
\hline
\multirow{2}{*}{TO} & pink & $15.72$ & $23.82$ & Ê$8.10$\\
& white & $25.84$ & $30.13$ & $4.29$\\
\hline
\multirow{2}{*}{NI} & pink & $20.94$ & $20.94$ & $0.00$\\
& white & $11.90$ & $17.38$ Ê& $5.48$\\
\hline
\multirow{2}{*}{MO} & pink & $5.72$ & $17.38$ & $11.66$\\
& white & $13.37$ & $14.13$ & Ê$0.76$\\
\hline
\multirow{2}{*}{MA} & pink & $13.37$ & $17.38$ & Ê$4.01$\\
& white & $14.13$ & $19.12$ & Ê$4.99$\\
\hline
\multirow{2}{*}{CH} & pink & $3.39$ & $12.62$ & $9.23$\\
& white & $12.62$ & $14.13$ & $1.51$\\
\hline
\multirow{2}{*}{CA} & pink & $35.98$ & $31.25$ Ê& $-4.73$\\
& white & $20.94$ & $32.4$ & $11.46$ \\
\hline
\end{tabular}
\end{threeparttable}
\end{center}
\vspace{-0.2cm}
\end{table}

\newpage
\section*{Figures}

\end{document}